\documentclass[aps,showpacs,a4paper]{revtex4}
\usepackage{graphicx}
\usepackage{amsmath}
\usepackage{dcolumn}

\def\ket{\rangle}
\def\<{\langle}
\def\>{\rangle}

\begin{document}
\preprint{APS/123-QED}
\title{Prime Factorization in the Duality Computer}

\author{ Wan-Ying Wang$^1$, Bin Shang$^2$, Chuan Wang$^1$ and
Gui Lu Long$^{1,3}$ }
\address{$^1$ Key Laboratory For Quantum
Information and Measurements and Department of Physics, Tsinghua University, Beijing
$100084$, People's Republic of China\\
$^2$ School of Computer Science \& Technology, Beijing Institute of
Technology, Beijing 100081, P. R. China\\
$^3$ Key Laboratory of Atomic and Molecular NanoSciences, Tsinghua\\
University, Beijing 100084, People's Republic of China }

\date{\today}

\begin{abstract}
    We give algorithms to factorize large integers
in the duality computer. We provide three duality algorithms for factorization based on
a naive factorization method, the Shor algorithm in quantum computing, and the Fermat's
method in classical computing. All these algorithms are polynomial in the input size.
\end{abstract}

\maketitle

\section{Introduction}\label{s1}
    Factorizing large integers has been a great problem in mathematics and has
    important applications in information security. Many efforts have been made to
    find polynomial algorithms.  The most naive method is that we
divide the input state $N$ by each integer ranging from 2 to
$\sqrt{N}$, so it needs $\sqrt{N}$ operations which needs
$2^{\frac{1}{2}\log N}$ steps, an exponential function of the input
size $\log N$. Until now, there has been no known efficient
polynomial algorithm for factoring a large integer in classical
computing. The most efficient factoring algorithms in classical
computing are the Fermat's factoring algorithm\cite{x1}.

With the advent of quantum computer\cite{feynman,benioff}, Shor gave a remarkable
 polynomial-time quantum algorithm for the factoring problem, and its computational
complexity has been proved to be $O((\log N)2(\log \log N)(\log \log \log
N))$\cite{1994}. Recently, a new computing model, the duality computer(DC), was
proposed based on the general principle of quantum interference\cite{r1}. The DC
exploits the particle wave duality properties of the quantum world, and provides
polynomial algorithms for NP-complete problems. Finding a marked item from an unsorted
database requires only a single step in the duality computer, which achieves the holy
grail in the unsorted database problem. Recently, the mathematical theory of the
duality computer has been worked out \cite{r2,r3}. In the DC, there are quantum wave
divider and combiner operations in addition to the usual unitary quantum gate
operations. The divider and combiner operators make the connection between duality
computing and classical computing very close, and it is possible that many classical
algorithms can be adopted to the duality computing with exponential speedup. In this
paper, we explore this connection with the problem of prime factorization as the
subject.

   This paper is organized as follows. In Sec.\ref{s2}, we briefly review
the DC and the implementation of the naive factorization algorithm.  In Sec.\ref{s3},
we briefly review Shor's algorithm and its implementation in the DC. Compared to the
quantum computer, the algorithm in the DC acquires some improvement though  both
algorithms are polynomial ones. In Sec.\ref{s4}, we design two algorithms using DC
based on both the Fermat's  methods and analyze their complexities. The results are
discussed and summarized in Sec.\ref{s5}.

\section{The Naive Factorization DC Algorithm }\label{s2}
    The general principle of quantum
interference for the quantum system was proposed and based on it, the DC was proposed
\cite{r1}. The DC makes use of not only the particle nature but also the wave nature of
microscopic objects. The quantum wave of the DC are split into several sub-waves by the
quantum wave divider(QWD) and they pass through different routes where different
computing gate operations are performed. These sub-waves are then re-combined to
interfere to give the computational results at the quantum wave combiner(QWC).

   The naive factorization algorithm can be easily implemented using the Long algorithm
   for unsorted database search \cite{r1}. We need a two register DC. The DC
   algorithm is given as follows

Step 0: foundlist=[null]

Step 1:  Prepare the the state of the first DC register in the equally distributed
state and the second register in the 0 state,
\begin{equation}
|\varphi_0\rangle=
\sqrt{\frac{1}{[\sqrt{N}]}}(|2\rangle+|3\rangle+\cdots+|[\sqrt{N}]+1\rangle)|0\ket,
\end{equation}
where $[ k ]$ means the integer nearest to $k$.

  Step 2: Let the DC go
through QWD, so that it divides the wave into two sub-waves,
\begin{equation}
|\varphi_u\rangle= \frac{1}{2}\sqrt{\frac{1}{[\sqrt{N}]}}(|2\rangle+|3\rangle+\cdots
+|p_i\rangle+\cdots+|[\sqrt{N}]+1\rangle)|0\ket,
\end{equation}
\[
|\varphi_d\rangle= \frac{1}{2}\sqrt{\frac{1}{[\sqrt{N}]}}(|2\rangle+|3\rangle+\cdots
+|p_i\rangle+\cdots+|[\sqrt{N}]+1\rangle)|0\ket.
\]

Step 3: Perform the following function to the lower-path sub-wave, \begin{eqnarray}
f(N,i)=\left\{\begin{array}{ll} 1 & {\rm if\; N\; mod\; i=0}\\
                                0 & {\rm otherwise}\end{array}\right.
\end{eqnarray}
Then the state of the lower-path becomes
\[
|\varphi_d'\rangle= \frac{1}{2}\sqrt{\frac{1}{[\sqrt{N}]}}
(|2\rangle|0\ket+|3\rangle|0\ket+\cdots+|p_i\rangle|1\ket+\cdots+|[\sqrt{N}]+1\rangle)|0\ket).
\]

Step 4: Apply a query to the lower-path wave: if the second register of the item is 1
retain the sign of the item, reverse the sign of other items and the items in the
foundlist.

Step 5: Recombine the two sub-waves $|\varphi_u\rangle$ and $|\varphi_d'\rangle$ at the
QWC. Then $|\varphi_f\rangle=|p_1\rangle+|p_2\ket+\cdots +|p_m\ket$ apart from a
normalization constant, namely $|\varphi_f\rangle$ contains all the factors except
those in the foundlist.

Step 6: Make a read-out measurement, and one of the unfound prime factors $p_i$ is
found out with probability 1. If nothing is found, then go to step 8.

Step 7: Add $p_i$ to foundlist.

Step 8: Stop.

If $N=p q$, then the above algorithm runs from step 0 to step 7 to give the smaller
factor, and then it goes to step 0 and flows to step 6, and then goes to step 8 and
stops.

\section{The DC Factorization Algorithm Based on the Shor algorithm}\label{s3}

The Shor algorithm finds the period of the function $f_{a,N}=a^x \;mod\; N$. In the
quantum computer, the registers of the quantum computer is divided into two, and the
Shor algorithm contains the following steps:

Step 1: Prepare the first register in the evenly distributed state and the second
register in state 0 \begin{eqnarray} |\psi_1\ket=\sqrt{1\over
q}\sum_{i=0}^{q-1}|i\ket|0\ket, \end{eqnarray} where $N^2<q \le 2N^2$.

Step 2: Apply the function $f_{a,N}$ to the state and store the results in the second
register, \begin{eqnarray} |\psi_2\ket=\sqrt{1\over q}\sum_{i=0}^{q-1}|i\ket|a^x\;mod
\;N\ket.
\end{eqnarray}

Step 3: Apply a Fourier transform to the first register.

Step 4: Makes a measurement on the first register and an integer will be found.

Repeat steps 1 to 4 a sufficient number of times and from the measured results  a
period is inferred, from which the period of the function $f_{a,N}(x)$ is determined.

In the DC there is no need to apply the  Fourier transform on the first register. Here
we present a modified Shor algorithm in the DC with $N=21$ and $a=2$ as an example.
Taking $q=512$, after applying the function $f_{a,N}$ on the DC, we let wave goes
through the QWD, so that its wave function is split into two parts, and in the lower
path we reverse the sign of all items except those whose second register number is 1
and the first register is not zero, namely the upper and lower path wave functions are
now
\begin{equation}
\begin{split}
|\varphi_u\rangle=\frac{1}{2}\frac{1}{\sqrt{512}}(&|0\rangle|1\rangle+|1\rangle|2\rangle+
|2\rangle|4\rangle+|3\rangle|8\rangle+|4\rangle|16\rangle+|5\rangle|11\rangle+\\
&|6\rangle|1\rangle+|7\rangle|2\rangle+|8\rangle|4\rangle+|9\rangle|8\rangle+|10\rangle|16\rangle+|11\rangle|11\rangle+\\
&|12\rangle|1\rangle+ \cdots),
\end{split}
\end{equation}
\begin{equation}
\begin{split}
|\varphi_d\rangle=\frac{1}{2}\frac{1}{\sqrt{512}}(&-|0\rangle|1\rangle-|1\rangle|2\rangle-
|2\rangle|4\rangle-|3\rangle|8\rangle-|4\rangle|16\rangle-|5\rangle|11\rangle+\\
&|6\rangle|1\rangle-|7\rangle|2\rangle-|8\rangle|4\rangle-|9\rangle|8\rangle-|10\rangle|16\rangle-|11\rangle|11\rangle+\\
&|12\rangle|1\rangle-\cdots ),
\end{split}
\end{equation}
respectively.

Next step, we combine the sub-waves at the QWC, and apart from an normalization
constant the wave becomes
\begin{eqnarray}
|\varphi_f\rangle&=&(|6\rangle|1\rangle+|12\rangle|1\rangle+\cdots+|504\rangle|1\rangle+|510\rangle|1\rangle
)\nonumber\\
&=&(|6\rangle+|12\rangle+\cdots+|504\rangle+|510\rangle)|1\rangle.
\end{eqnarray}

Make a read-out measurement of the first register, one will obtain one of the following
numbers: 6, 12, 18, 24 ... 504, 510 with equal probability. Repeat the procedure a
number of times, the smallest increment will be found and hence the required period.

\section{DC Factorization Algorithm based on Fermat's method}
\label{s4}
 On classical computers, Fermat's method\cite{x1} for factoring an
odd integer $N=pq$ where $q,$ and $p$ are prime integers consists of finding positive
integers $X$ and $Y$ that satisfies $N=x^2-y^2$. The key points lie in finding those
$X$ and $Y$ satisfying $N=X^2-Y^2$ so that we can get the result $p=(X+Y)/2,q=(X-Y)/2$.
Fermat's method is quite efficient if $p/q$ is near 1, but less efficient if $p/q$ is
not near 1. The computational complexity for Fermat's method is $\mathcal
{O}(q-\sqrt{N})$, and in worst conditions is $\mathcal {O}(N/2-\sqrt{N})$

We design an algorithm under worst conditions for a DC based on Fermat's method and
analyze its complexity. We denote the integer bigger than or equal to $(N/2-\sqrt{N})$
as $M$
\begin{enumerate}
\item Make the initial state
\begin{equation}
|\varphi_0\rangle=\frac{1}{\sqrt{M}}\sum_{i=0}^{M}|x\rangle
=\frac{1}{\sqrt{M}}(|x_0\rangle+|x_1\rangle+\cdots+|x_{M}\rangle)
=\frac{1}{\sqrt{M}}(|\sqrt{N}\rangle+|\sqrt{N}+1\rangle+\cdots+|N/2\rangle)
\end{equation}

\item Let $|\varphi_0\rangle$ go through the QWD and it is
divided into:
\begin{equation}
|\varphi_u\rangle=\frac{1}{2\sqrt{M}}\sum_{i=0}^{N/2}|x_i\rangle
=\frac{1}{2\sqrt{M}}(|x_0\rangle+|x_1\rangle+\cdots+|x_{M}\rangle)
=\frac{1}{2\sqrt{M}}(|\sqrt{N}\rangle+|\sqrt{N}+1\rangle+\cdots+|N/2\rangle)
\end{equation}
\begin{equation}
|\varphi_d\rangle=\frac{1}{2\sqrt{M}}\sum_{i=0}^{M}|x_i\rangle
=\frac{1}{2\sqrt{M}}(|x_0\rangle+|x_1\rangle+\cdots+|x_{M}\rangle)
=\frac{1}{2\sqrt{M}}(|\sqrt{N}\rangle+|\sqrt{N}+1\rangle+\cdots+|N/2\rangle)
\end{equation}

\item Apply on the upper-sub wave the following query function $f_{x,N}$
\begin{eqnarray}
f_{x,N}=\left\{\begin{array}{ll} +1 & {\rm if}\; \sqrt{x^2-N} {\rm \;is\;} {\rm an\; integer}\\
                                 -1 & {\rm otherwise}\end{array}
                                 \right.
                                 \end{eqnarray}

\item Let the two sub-waves go through the QWC and it is combined into $|X\rangle$,
then we can make a measurement to read out this $X$, and the $Y$ is then
$\sqrt{X^2-N}$. Then $p$ and $q$ are obtained.
\end{enumerate}

The complexity for our algorithm is $\mathcal {O}(\log{(N/2}))$

\section{Summary}\label{s5}
   In this paper, we propose three duality
algorithms to factorize large integer number.  It has been shown that Shor's algorithm
in a duality computer can be simplified a little. The naive factorization algorithm can
be implemented in a duality computer polynomially. We also show that the Fermat's
method in classical computing can be adopted in duality computing with a computational
complexity of $O(\log N)$.

  Compared to quantum computing, the computational complexity of prime-factorization
  algorithm is the same which is not remarkable. The important point shown in this work
  is that exponential algorithms in classical computing can be implemented in duality
  computer with a polynomial computational complexity. This provides a convenient
  avenue in constructing duality computing algorithms.

\begin{acknowledgments}
This work is supported by the National Fundamental Research Program Grant No.
001CB309308, China National Natural Science Foundation Grant No. 10325521, 60433050,
and the SRFDP program of Education Ministry of China.
\end{acknowledgments}

\end{document}